# Surrogate Modeling of Fluid Dynamics with a Multigrid Inspired Neural Network Architecture

Quang Tuyen Le[1], Chinchun Ooi[1]

[1]Department of Fluid Dynamics, Institute of High Performance Computing, 1 Fusionopolis Way, #16-16 Connexis, Singapore 138632, Singapore.

Author e-mail address: leqt@ihpc.a-star.edu.sg

Corresponding author e-mail address: ooicc@ihpc.a-star.edu.sg

Corresponding author contact number: +65 64191503

Corresponding author fax number: +65 64670200

**Abstract**

Algebraic or geometric multigrid methods are commonly used in numerical solvers as they are a multi-resolution method able to handle problems with multiple scales. In this work, we propose a modification to the commonly-used U-Net neural network architecture that is inspired by the principles of multigrid methods, referred to here as U-Net-MG. We then demonstrate that this proposed U-Net-MG architecture can successfully reduce the test prediction errors relative to the conventional U-Net architecture when modeling a set of fluid dynamic problems. In total, we demonstrate an improvement in the prediction of velocity and pressure fields for the canonical fluid dynamics cases of flow past a stationary cylinder, flow past 2 cylinders in out-of-phase motion, and flow past an oscillating airfoil in both the propulsion and energy harvesting modes. In general, while both the U-Net and U-Net-MG models can model the systems well with test RMSEs of less than 1%, the use of the U-Net-MG architecture can further reduce RMSEs by between 20% and 70%.

## 1. Introduction

Numerical simulations are extremely useful in many engineering applications, as they can be both more economical and time effective than actual experimental testing, especially when parametric studies are desired in the design iteration phase. Computational Fluid Dynamics (CFD) in particular, has shown great utility as a design tool across many industries such as aerospace and marine and offshore.

In recent years, neural networks have been shown to be a universal function approximator, and have found widespread application in computer vision and classification of nonlinear systems (Hornik, Stinchcombe, & White, 1989). More interestingly, it has also found great utility in the modeling of engineering systems such as in fluid dynamics. Neural networks have been shown to complement numerical models in areas such as turbulence modeling, force prediction and flow reconstruction (Dao,

Nguyen, Ooi, & Le, 2020; Holland, Baeder, & Duraisamy, 2019; Ling, Kurzawski, & Templeton, 2016; Muralidhar et al., 2020; Ooi et al., 2020; Parish & Duraisamy, 2016; Umetani & Bickel, 2018; Ye et al., 2020). In particular, the convolutional neural network (CNN) has been shown to be effective for prediction across different flow scenarios while providing flexibility in the treatment of variable geometries. For example, CNNs have been shown to effectively predict important aerodynamic quantities of interest such as pressure distribution, lift and drag across different airfoil shapes and flow conditions such as Reynolds number (Re) (Yilmaz & German, 2017; Zhang, Sung, & Mavris, 2018), while other authors further moved beyond single metrics or quantities of interest by demonstrating prediction of full flow fields for various airfoil shapes across multiple flow regimes (Bhatnagar, Afshar, Pan, Duraisamy, & Kaushik, 2019; Chen, Viquerat, & Hachem, 2019; Sekar, Jiang, Shu, & Khoo, 2019). In addition to the modeling of steady-state flow scenarios, CNNs can also be adopted for unsteady flows. While there has been work suggesting that CNNs can out-perform recurrent neural networks on sequence data (Bai, Kolter, & Koltun, 2018), Long Short Term Memory (LSTM) models or other recurrent neural networks have typically been proposed as good surrogate models for capturing the temporal dynamics of unsteady aerodynamics (Li, Kou, & Zhang, 2019). Hence, hybrid models combining CNNs and LSTM-type models have been proposed, whereby the convolutional component is used for feature extraction and dimensionality reduction while the LSTM models encode dynamical information, and these models have been shown to be effective at capturing the spatial-temporal features of various flows (Han, Wang, Zhang, & Chen, 2019; Hasegawa, Fukami, Murata, & Fukagata, 2020). In particular, we note that CNNs are very advantageous for feature extraction, and have shown very good performance when applied to problems such as flow reconstruction in fluid dynamical systems. For example, Fukami et al. showed that CNNs and its variant can be very effective for super-resolution reconstruction of a turbulent flow field over cylinder, while other authors have similarly applied CNNs to purposes such as data recovery and flow field reconstruction (de Frahan & Grout, 2019; Fukami, Fukagata, & Taira, 2019; Jin, Cheng, Chen, & Li, 2018; Lee & You, 2019).

While these different machine learning models have been shown to be useful, the majority of current work utilize machine learning models or neural networks as pseudo-black boxes, with little consideration of the mechanism by which these models are learning or representing the physics-based problem. Hence, recent work in the area of physics-guided machine learning is particularly interesting for showing how incorporating domain knowledge into the selection or design of model architecture can improve these models (Jia et al., 2020; Karpatne et al., 2017; Raissi, Perdikaris, & Karniadakis, 2019; Zhu, Zabaras, Koutsourelakis, & Perdikaris, 2019).

Algebraic multigrid methods are commonly integrated into numerical solvers used in scientific computing and are known to be very effective for problems with multiple scales, including in fluid dynamics (Lonsdale, 1993; Ruge & Stüben, 1987; Webster, 1994; Weiss, Maruszewski, & Smith, 1999). Fundamental to their good performance is the idea that a multigrid hierarchy allows for information transfer using local procedures at different length scales, and ideas from multigrid methods have been incorporated into some particularly well-performing neural networks, such as the U-Net architecture (Milletari, Navab, & Ahmadi, 2016; Ronneberger, Fischer, & Brox, 2015). The idea of integrating multigrid architecture into neural networks has thus been shown to provide enhanced performance relative to conventional CNN architecture in computer vision problems (Ephrath, Ruthotto, & Treister, 2020; He & Xu, 2019; Ke, Maire, & Yu, 2017), with He et al. noting the close correspondence between algebraic multigrid methods and multigrid neural networks, while others have demonstrated advantages with multilevel or multiscale neural networks (Haber, Ruthotto, Holtham, & Jun, 2017; Pelt & Sethian, 2018; Scott & Mjolsness, 2019). Essentially, the communication of spatial information is improved relative to standard convolutional neural networks by the concatenation of neighboring grid scales. Hence, a multigrid-based neural network can better facilitate cross-scale information exchange and improve the network's representational abilities. Ke et al. in particular demonstrated that such scale-space routing of information is a good alternative to typical attempts to improve CNN performance by going wider or deeper, with improved performance relative to baseline CNN models (Ke et al., 2017).

Based on the utility of multigrid methods in CFD, we thus further hypothesize that the integration of a multigrid structure into a CNN architecture can facilitate and improve the model's representation of engineering systems. In particular, the widely-used U-Net convolutional neural network has been shown to be very effective across multiple domains including fluid dynamics, and is strikingly similar in structure to the conventional multigrid solver's V-cycles (Ronneberger et al., 2015; Thuerey, Weißenow, Prantl, & Hu, 2020). Hence, we evaluate the U-Net model and a multigrid-inspired U-Net-MG model for their ability to learn and reproduce flow fields.

In the following sections, we present our results on the following fluid dynamical systems: 1) flow past a stationary cylinder, and flow due to two cylinders in out-of-phase motion; and 2) flow past an oscillating foil. These examples are chosen as the former set of scenarios are commonly studied canonical problems in fluid dynamics, with relevance to the marine industry such as in understanding the interaction between risers (Hu & Zhou, 2008; Huera-Huarte & Jiménez-González, 2019; Liu & Jaiman, 2016; Ooi et al., 2020; Wang, Xiao, Zhu, & Incecik, 2017), while the latter set of scenarios are greatly relevant to the areas of propulsion or energy harvesting (Andersen, Bohr, Schnipper, & Walther, 2017; Jihoon Kim et al., 2015; Kinsey & Dumas, 2008; Le & Ko, 2015; Read, Hover, & Triantafyllou, 2003). Critically, they also encompass an interesting range of problems in surrogate modeling, including flow in a scenario with a fixed object and constant boundary conditions (flow past a stationary cylinder), flow in a scenario with varying inlet boundary conditions and moving objects (flow past two cylinders in out-of-phase motion), and flow in a scenario with consistent inlet boundary conditions and moving objects but with varying geometries (flow past oscillating foils of different geometries).

## 2. Methods

### 2.1. U-Net Architecture

Convolutional neural networks are commonly used for handling images (grid-like) data in fields like computer vision as the individual convolutional layers have been shown to be effective at feature extraction. Hence, as a baseline, we adopt the U-Net convolutional neural network architecture which has been demonstrated to be extremely suited to many tasks such as image classification in biomedical systems and in the modeling of fluid dynamics systems (Chen et al., 2019; Falk et al., 2019; Ronneberger et al., 2015; Thuerey et al., 2020). In particular, Thuerey et. al. proposed the use of a U-Net model and demonstrated very good flow and pressure field predictions across different airfoils and flow conditions with test errors of under 3% (Thuerey et al., 2020).

Briefly, the U-Net has a bowtie structure, comprising separate encoder and decoder halves. The encoder half extracts features via successive layers of convolutional kernels while the decoder half converts the extracted lower-dimensional features back into the desired output variable with a pre-determined spatial resolution. In addition, a skip-connection is implemented to connect each corresponding branch of the encoder and decoder halves of a U-Net in order to enhance the propagation of information through the multiple layers of the neural network.

For all U-Net models used in this study, the input variables are provided as a set of 2-D arrays, which are down-sampled by a constant factor of 2 through each layer with a convolutional kernel of size 4 × 4. The number of feature channels is also progressively increased by a factor of 2 at each stage to allow the network to extract increasingly large-scale and abstract information. Effectively, each stage can thus be thought of as comprising a convolutional layer, a batch normalization layer and a non-linear activation component, with the choice of activation function being the leaky ReLU ($\alpha = 0.2$) in the encoder half. This is reversed in the decoder half of the U-Net, whereby each layer is progressively up-sampled by a factor of 2, the number of feature channels are gradually reduced by a factor of 2, and the ReLU function is used as the non-linear activation function. A schematic of the U-Net architecture used is provided in Fig. 1, with additional information such as the number of filters at each stage and the corresponding Keras functions demarcated in the legend.

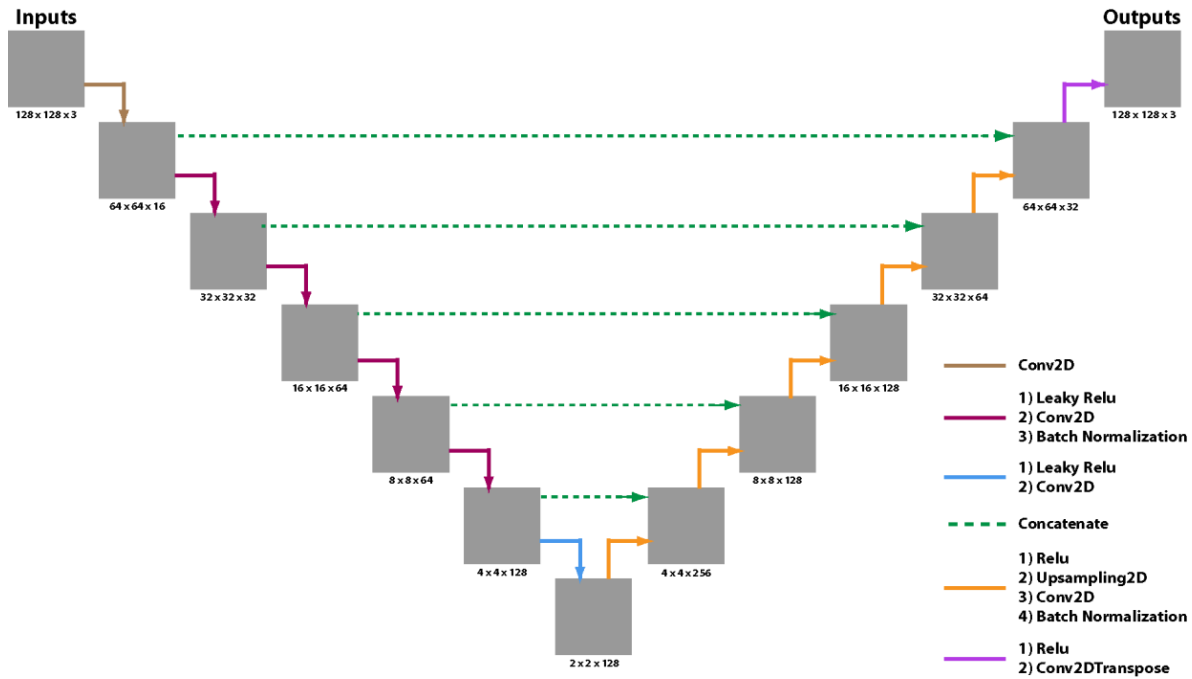

**Fig. 1.** Schematic of the U-Net architecture used in this work as a baseline for comparison. The different operations in each step are represented by the differently-colored arrows in the schematic. Operations are repeated for each successive layer in the encoder and decoder halves, except for the very top and very bottom of the U-Net.

### 2.2. Multigrid-inspired U-Net Architecture (U-Net-MG)

As per work by Ke et al. (Ke et al., 2017), we replace the individual grids in each layer of the U-Net architecture with 3 grids of varying resolutions in our proposed U-Net-MG architecture. Hence, for each successive layer of the encoder half of the neural network, the convolution is able to work on 3 grids of varying spatial scales. The outputs from applying the convolution on each grid in each layer are subsequently concatenated to serve as an input to the next layer.

In this study, we chose to use Max-pooling and Sub-sampling operations as lateral communication tools between grids of three different resolutions at each stage as depicted in Fig. 2, as this would avoid the introduction of additional parameters that would need to be further optimized during training. In addition, this bears direct similarity to the simplest prolongation and restriction operators commonly used in algebraic multigrid solvers such as arithmetic averaging.

In order to facilitate comparison, we used an identical number of convolutional kernels and network architecture hyperparameters such as number of up-sampling and down-sampling stages in both the original U-Net and the proposed U-Net-MG in this work. As the number of convolutional kernels is the primary determinant of the number of parameters in each neural network, this also ensures that the U-Net-MG architecture does not significantly increase the number of parameters to be trained. A detailed schematic of the implemented U-Net-MG architecture is provided in Fig. 2 with additional information such as the number of filters at each stage and the corresponding Keras functions used for easy reference.

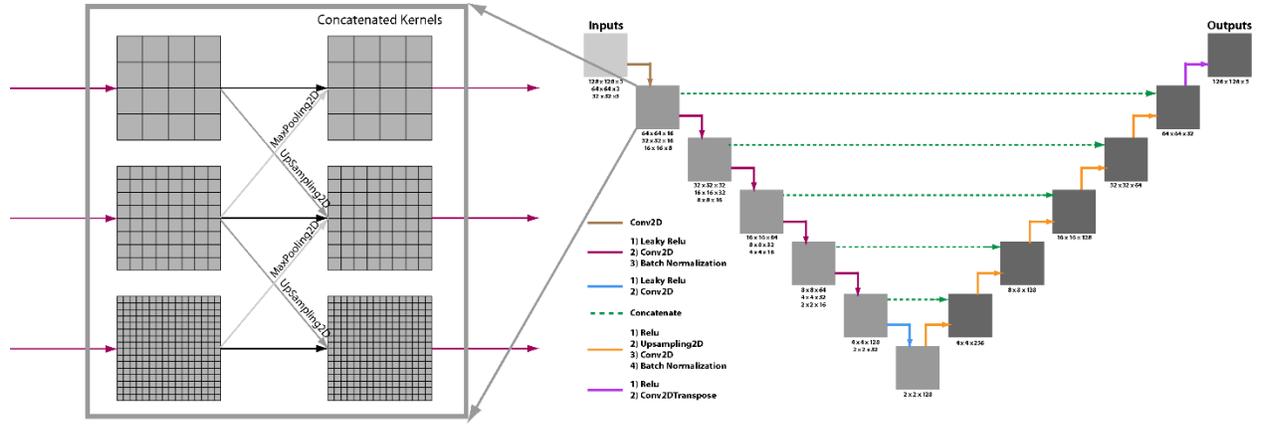

**Fig. 2.** Schematic of the proposed U-Net-MG architecture used in this work. In addition, a schematic is provided in the inset depicting how 3 grids of varying resolutions are used in the encoder half of the U-Net to further facilitate propagation of information across scales within the neural network. The number of channels and layers are maintained with the U-Net architecture across each problem to ensure consistency in comparison.

### 2.3. Data Pre-processing and Model Training

Default network hyper-parameters were chosen based on prior work by Thuerey et al. which similarly utilized a U-Net architecture on fluid dynamics problems and had reported good predictive performance (Thuerey et al., 2020). Standard Keras implementations of the ADAM optimizer and required activation functions and convolutional operations were used. All cases were trained for $10^4$ epochs, as it was empirically observed that all cases appeared to reach a minimum in training error within $10^4$ epochs. In general, a 9:1 training and test split of the data-set was used across the different flow scenarios evaluated. Mean squared error was the loss function used for training via the backpropagation algorithm as defined by:

$$MSE = \frac{1}{N}\sum_{m=1}^{N}\sum_{X}\left(F_{X,m}^{true} - F_{X,m}^{pred}\right)^2 \quad (1)$$

where N is the number of samples, X is the index corresponding to each pixel of the flow domain, F is one of u-velocity, v-velocity or pressure within the domain, and the superscripts *true* and *pred* refer to the true value as obtained from CFD simulations, and as predicted by the neural network respectively.

In accordance with prior work, the following pre-processing of data is implemented: 1) normalization of all involved parameters with respect to free-stream condition ( $\tilde{U} = U/U_\infty$, $\tilde{p} = p/U_\infty^2$ ), and 2) subtraction of the mean pressure value so as to standardize the reference pressure point across different scenarios ($\hat{p} = \tilde{p} - p_{mean}$) (Thuerey et al., 2020).

## 3. Results

### 3.1. CFD Simulation Results

In the following sub-sections, three sets of fluid dynamics simulations are described. These cases are typical fluid dynamics problems commonly studied in literature, and are used here as a basis for comparing the performance of the U-Net model and the proposed U-Net-MG model (Jihoon Kim et al., 2015; Mahir & Rockwell, 1996; Ooi et al., 2020; Read et al., 2003). All simulations are run via the open-source OpenFOAM platform with the k-ω SST turbulence model. Further details on domain and solver settings are provided in the individual scenario sub-sections.

#### 3.1.1. Flow over Single Stationary Cylinder

In the first case, the 2D scenario of turbulent flow past a stationary cylinder is modeled. The full CFD simulation is run with the pimpleFoam solver in OpenFOAM. The full domain comprises of 54300 hexahedral cells, with the inlet and outlet at distances of 13.5D and 50D, where D is the diameter of the cylinder, as depicted in Fig. 3. In this flow scenario, the cylinder diameter is of value D = 1 m, while the inlet velocity boundary condition is a constant U = 1 m/s. A non-slip boundary condition for velocity is applied on the cylinder surface. A kinematic viscosity ν, of $8.695 \times 10^{-6}$ m²/s is used in this simulation, which corresponds to a Reynolds number of $1.15 \times 10^5$.

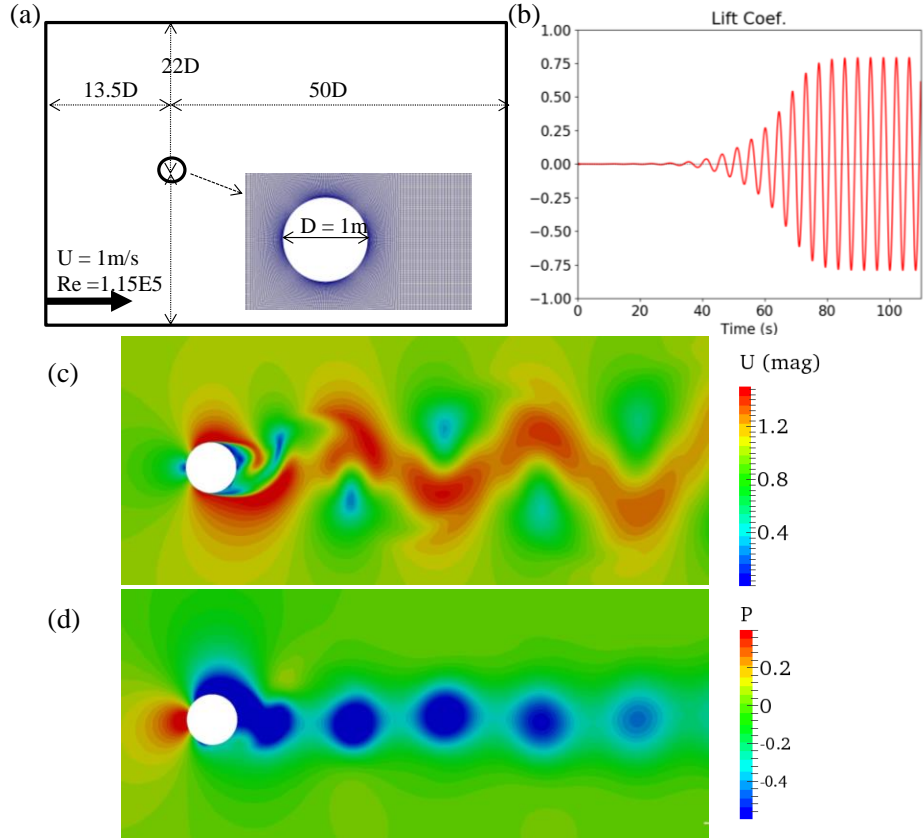

**Fig. 3.** (a) Schematic of the simulation domain is provided, depicting the domain dimensions and boundary conditions. (b) Plot of the lift coefficient obtained for the stationary cylinder shows how the transient simulations approach a periodic state. (c)-(d) Representative contour plots of the (c) velocity magnitude and (d) pressure around the stationary cylinder.

The lift coefficient is also calculated for this transient simulation, and shows the gradual progression of the flow to its natural frequency, as depicted in Fig. 3, while representative velocity and pressure contour plots are provided to illustrate the appearance of a von Karman vortex street as expected. After the flow has stabilized, an additional 100 snapshots of velocity and pressure were recorded, and this data was subsequently used as data for modelling with our proposed neural networks. As the majority of the interesting flow behavior is centered around the cylinder, a reduced domain of range -0.75D < x < 3.25D and -1D < y < 1D is used, where the cylinder is centered at (x,y) = (0,0).

### 3.1.2. Two Cylinders in Out-of-phase Motion

Further expanding on the simulation of flow past a stationary cylinder, we then consider the case of two cylinders in plunge motion with opposing phase, also referred to as "out-of-phase" motion. The vertical position of each cylinder is described by the following equation:

$$h(t) = 0.5D sin(\omega t + \varphi) \tag{2}$$

where $\omega$ is the frequency of plunge motion, $\omega = 2\pi$ rad/s, and $\varphi$ is the phase shift between the two cylinders, $\varphi = \frac{\pi}{2}$ rad/s. The gap between the two cylinders is of length 4D while the domain size is maintained as per the flow past stationary cylinder scenario. As this simulation involves mesh movement and deformation, the pimpleDyMFoam solver from OpenFOAM is used, with a slightly larger mesh size of 56100 hexahedral cells. Other flow boundary conditions such as inlet boundary conditions and kinematic viscosity are maintained as per the flow past a stationary cylinder, thus yielding an identical Re of $1.15 \times 10^5$.

Sample velocity and pressure contour plots are provided in Fig. 4, further illustrating how this flow scenario is more complex due to the interaction between the two cylinders in plunge motion. It is worth noting that the two cylinders plunge at a frequency of 1 Hz which is higher than its natural frequency (~0.25 Hz). Therefore, the flow field is dominated by the cylinders' movement and is more complex than the first case of flow past a single, stationary cylinder. A set of 100 snapshots of the velocity and pressure field solutions are also collected for a single complete cycle after the flow has stabilized.

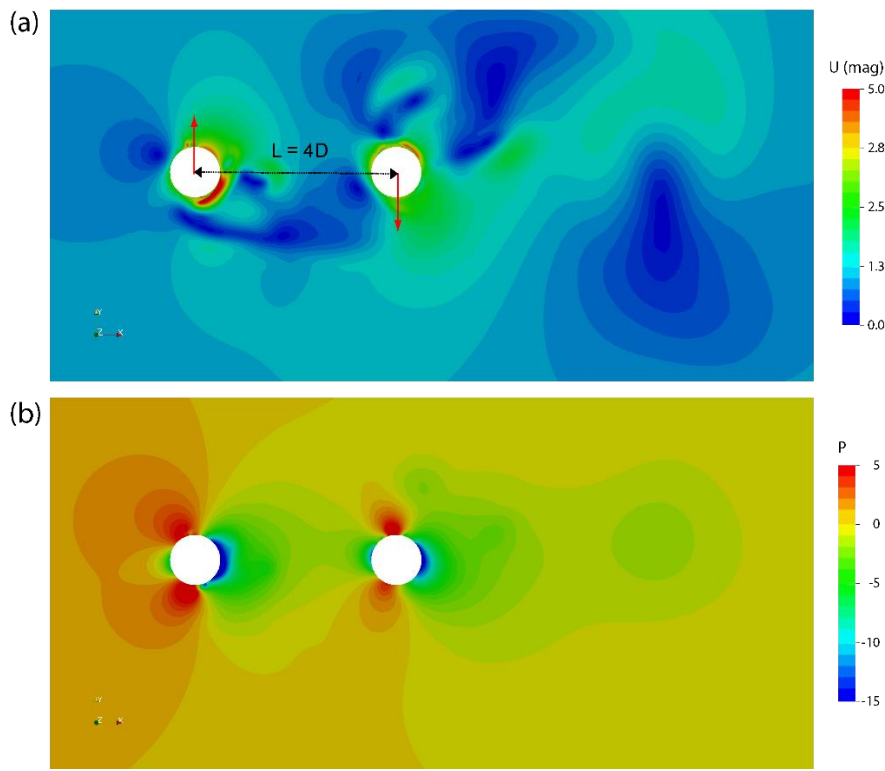

**Fig. 4.** Sample contour plots of (a) velocity magnitude and (b) pressure for two cylinders that are in out-of-phase motion.

In addition, the same set of simulations are run for an additional 3 inlet velocities, U = 0.5 m/s, U = 1.5 m/s, U = 2.0 m/s, for modeling with the neural networks. This was chosen to further evaluate the ability of the U-Net and U-Net-MG models to represent the flow field accurately across variations in inlet velocities and consequently, Reynolds number.

### 3.1.3. Hydrodynamics of Oscillating Foil

An additional interesting and more complicated use case is in the simulation of oscillating foils for their performance in propulsion or energy harvesting applications, such as in prior work by Read et al (Read et al., 2003). Similar to their study, we simulated the flow past an oscillating NACA 0012 foil in order to model its performance under both propulsion conditions and energy harvesting conditions. For this particular set of simulations, a chord length of c = 0.1 m is chosen with an incoming velocity of U = 0.4 m/s. Hence, the simulations correspond to a Re of $4 \times 10^4$ under a dynamic viscosity of $10^{-6}$ m²/s and standard water conditions. PimpleDyMFoam with diffusivity quadratic function is used to control mesh deformation while both translational and rotational motions are coupled. The full simulation domain comprises of 45000 hexahedral cells, with a horizontal and vertical length that is 50 and 40 times the chord length respectively.

The foil is specified to have the following heaving and pitching motions:

$$h(t) = H \sin(\omega t) \tag{3.1}$$

$$\theta(t) = \theta_0 \sin(\omega t + \varphi) \tag{3.2}$$

where the amplitude of heave motion is given by H = 0.75c and the circular frequency is defined as $\omega = \pi$ rad/s. This corresponds to a Strouhal number ($St = \frac{2fH}{U}$) of 0.1875. The phase angle between the pitch and heave motions is defined as $\varphi = \frac{\pi}{2}$ rad/s. The resulting angle of attack, $\alpha(t)$, is a function of heave velocity and pitch angle, which in turn determines the output of the oscillating foil.

$$\alpha(t) = -\arctan\left(\frac{\dot{h}(t)}{U}\right) + \theta(t) \tag{4}$$

When the angle of attack is small, the oscillating foil can produce a thrust, which is commonly referred to as propulsion mode in fish robots. On the other hand, for a large angle of attack, the foil can extract energy from the flow, as is commonly employed in tidal energy installations. Here, we choose to investigate both modes of operation by selecting two pitch angles, $\theta_0$, of amplitude 15° and 40° which would correspond to a propulsion mode and an energy harvesting mode respectively. A schematic of the NACA 0012 airfoil as simulated is displayed in Fig. 5, along with representative velocity and pressure contour plots of the simulation.

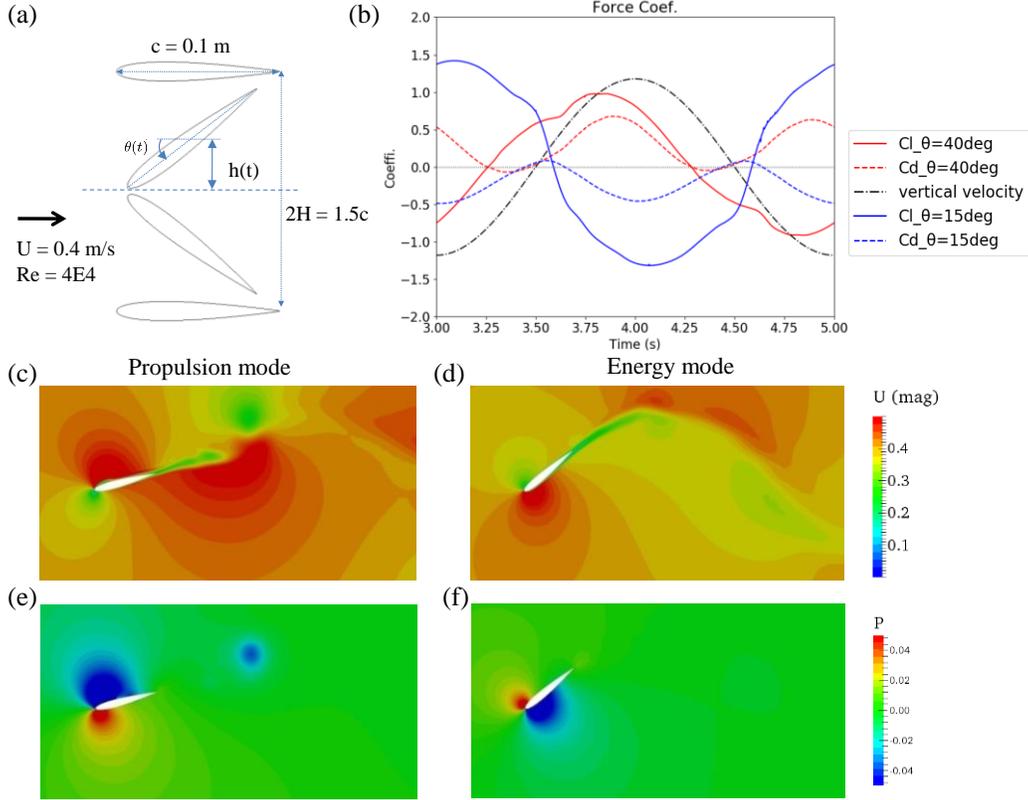

**Fig. 5.** (a) Schematic of the NACA 0012 used in this simulation. (b) is a plot of the lift and drag coefficients experienced by the foil for different modes of operation. (c) and (d) are velocity contour plots and (e) and (f) are pressure contour plots of the flow under propulsion ($\theta_0 = 15^o$) and energy harvesting ($\theta_0 = 40^o$) modes respectively.

It should be noted that the power from such oscillating foils derives from the plunge motion and positive power is generated only if the lift and vertical velocity are "in-phase". As illustrated in the graph of force coefficients under the pitch angles of 15º and 40º, thrust is generated for $\theta_0 = 15^o$ while drag is generated for $\theta_0 = 40^o$. Similar to the previous cases of flow past cylinders, snapshots of the velocity and pressure field are extracted after simulations have reached a steady state.

In addition, the same set of simulations are run for an additional 3 NACA foil geometries, NACA 0020, NACA 0025 and NACA 0030, under energy harvesting mode ($\theta_0 = 40^o$) for modeling with the neural networks. This was chosen to further evaluate the ability of the U-Net and U-Net-MG models to represent the flow field accurately across variations in geometry.

### 3.2. U-Net and U-Net-MG Results

#### 3.2.1. Flow over Single Stationary Cylinder

Each snapshot is recorded for network inference as an image with spatial resolution of $256 \times 128$. As a first proof-of-concept, 3 fields are provided as input to the neural networks: current U velocity, current V velocity and geometry, while the output field is the corresponding pressure. Fig. 6 shows an example of the input and output data provided to the U-Net and U-Net-MG models.

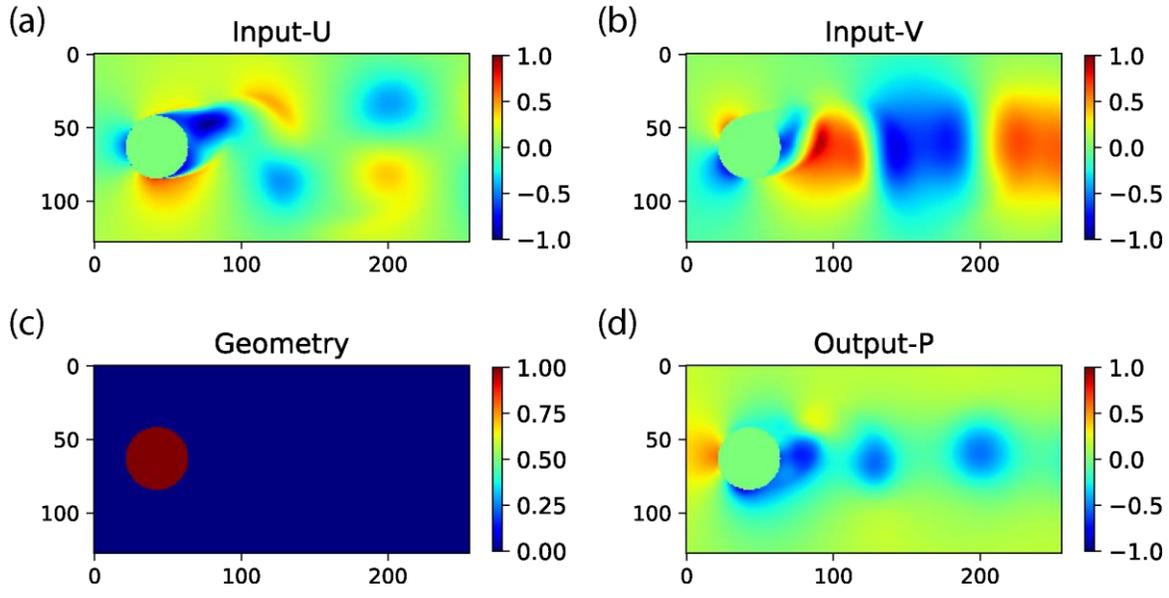

**Fig. 6.** Example of input and output fields provided to the neural network models for training. (a), (b) and (c) are input fields encoding U-velocity, V-velocity and a Geometry field, while (d) is an example of the pressure field as predicted by the neural network.

An example of the predicted pressure from both U-Net and U-Net-MG is shown in Fig. 7, in addition to the reference value obtained from CFD simulations. In this first example, both networks can reproduce the CFD solution for pressure reasonably well, with very similar average test RMSE of less than 0.21% for the U-Net model and 0.17% for the U-Net-MG model. Notably, the models can represent the region near the cylinder fairly well despite the large pressure jump. In this particular example, the use of a U-Net-MG model improved the test error relative to the base U-Net model by approximately 18%.

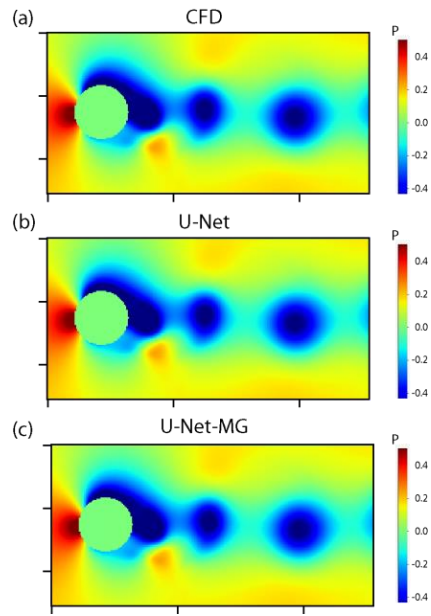

**Fig. 7.** Pressure contours of an example case for flow past a stationary cylinder are presented for (a) the CFD reference, (b) the U-Net model prediction, and (c) the U-Net-MG model prediction.

### 3.2.2. Two Cylinders in Out-of-phase Motion

In this next flow scenario, we assume that the velocity field information for the entire domain is also not available, although the motion of the individual cylinders and a far-field free-stream velocity is known. This could also correspond to a simple digital twin system, whereby single sensors are mounted on the solid cylinders of interest, and sensors are available to measure the free-stream velocity.

Hence, in this scenario, the networks are trained to predict the full U, V and p for the domain directly from simple input conditions (free-stream velocity boundary and velocity of each cylinder). The first and second input data channels are the boundary conditions for U and V for the far-field and for each cylinder as specified by Equation 1, while the third input channel remains the geometry field. The output data here is then U, V and p, across the domain of interest. Additional data pre-processing, and neural network hyper-parameters are maintained as per the flow past a stationary cylinder. Illustrative input and output fields are provided in Fig. 8.

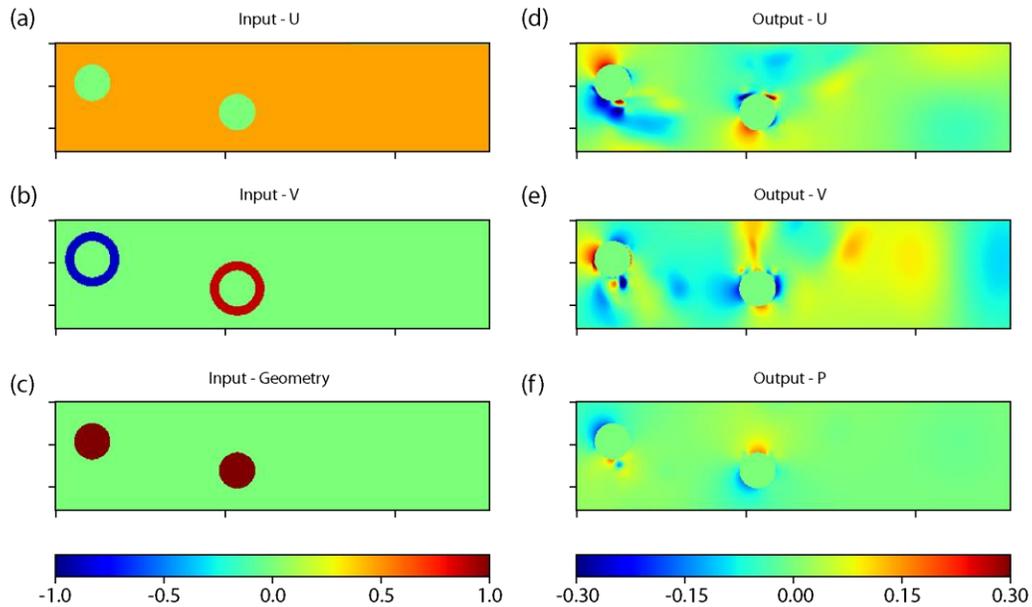

**Fig. 8.** Example of input and output fields provided to the neural network models for training for the case of flow past two cylinders in out-of-phase motion. (a), (b) and (c) are input fields encoding U-velocity, V-velocity and a Geometry field, while (d), (e) and (f) are an example of the U-velocity, V-velocity and pressure field as predicted by the neural network.

Fig. 9 shows sample velocity and pressure contours predicted from both U-Net and U-Net-MG models as well as the reference CFD solution. These contours show that both U-Net and U-Net-MG can reproduce the CFD result well for this particular flow scenario. The graphs of U, V, p through the two cylinders and downstream through the domain also illustrate how these networks can successfully reproduce the velocity and pressure.

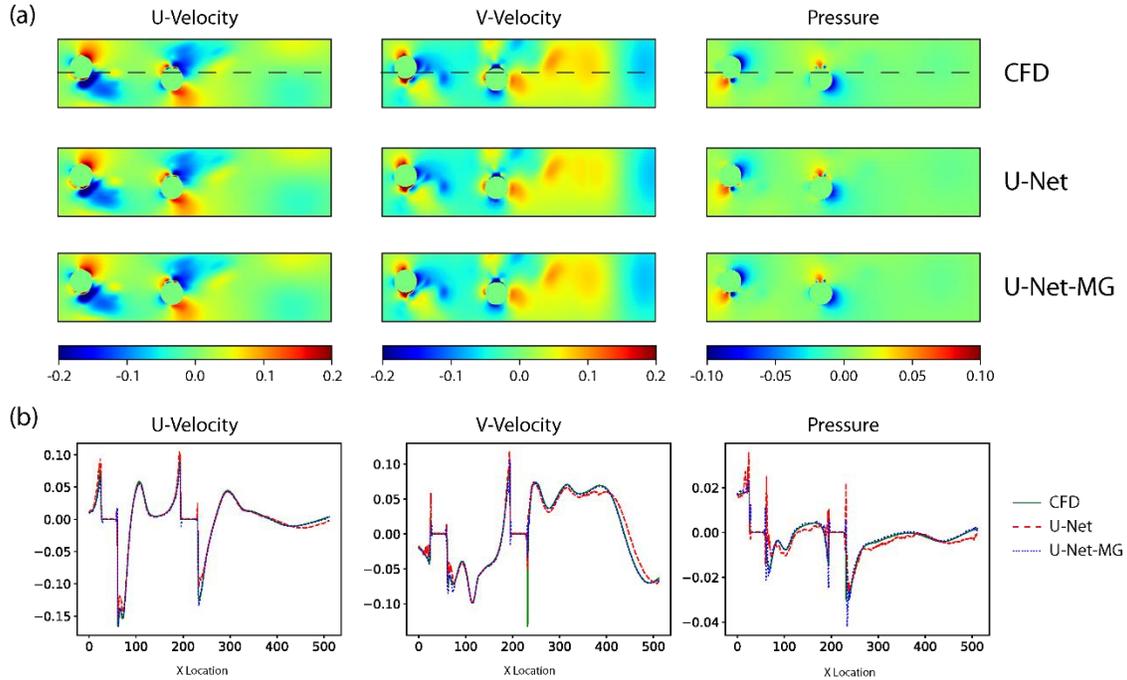

**Fig. 9.** (a) Sample contour plots derived from a CFD reference and predicted outputs from U-Net and U-Net-MG for U, V and P are provided for an example case. (b) The line graphs show the values of U, V and P through the cylinders and downstream of the cylinders along the dotted line depicted in (a).

The test RMSEs are between 0.1% and 1% for all flow variables across both the U-Net and U-Net-MG models, with the error in V being largest for both models, with errors of 0.98% for the U-Net model and 0.46% for the U-Net-MG model. This may be due to the larger gradients in the V field due to such high reduced frequency plunge motion. Nonetheless, we note that the U-Net-MG again produces better prediction accuracy than U-Net, with relative improvements of 37% for U, 53% for V, and 22% for pressure.

The test RMSEs for both the flow past stationary cylinder and the flow past two cylinders in out-of-phase motion are plotted in Fig. 10, and illustrate how the U-Net-MG model improves predictions relative to the U-Net model.

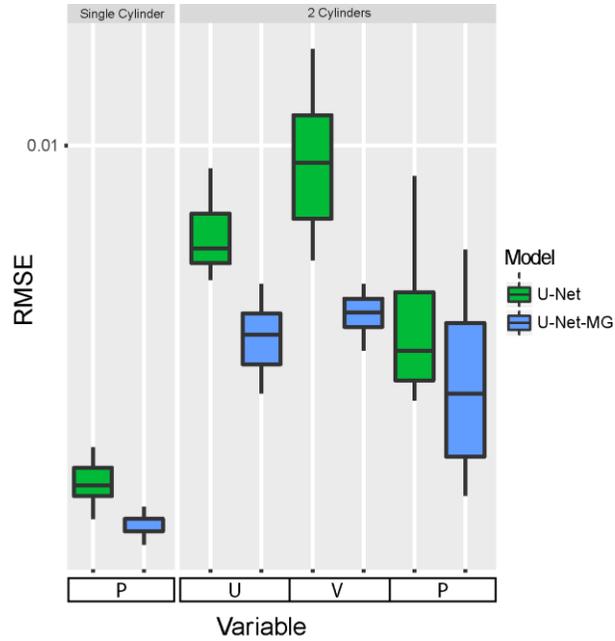

**Fig. 10.** Boxplot of the test RMSEs obtained for the U-Net and U-Net-MG models constructed for the case of flow past a stationary cylinder and case of two cylinders in out-of-phase motion.

In addition, the models are trained with a training set comprising snapshots from CFD simulations for $U = 0.5 \text{ or } 1.0 \text{ or } 2.0 \text{ } m/s$ and are tested with snapshots for $U = 0.5 \text{ or } 1.0 \text{ or } 1.5 \text{ or } 2.0 \text{ } m/s$. More specifically, the training set comprises 90 samples from U = 1.0 m/s, 75 samples from 0.5 m/s and 2.0 m/s, while the test set comprises 10 samples from U = 1.0 m/s, 25 samples from 0.5 m/s and 2.0 m/s, and 100 samples from U = 1.5 m/s. Hence, we are testing the ability of the U-Net and U-Net-MG models to approximate the flow for scenarios with U = 1.5 m/s that were not in the initial training dataset. Results are presented in Fig. 11.

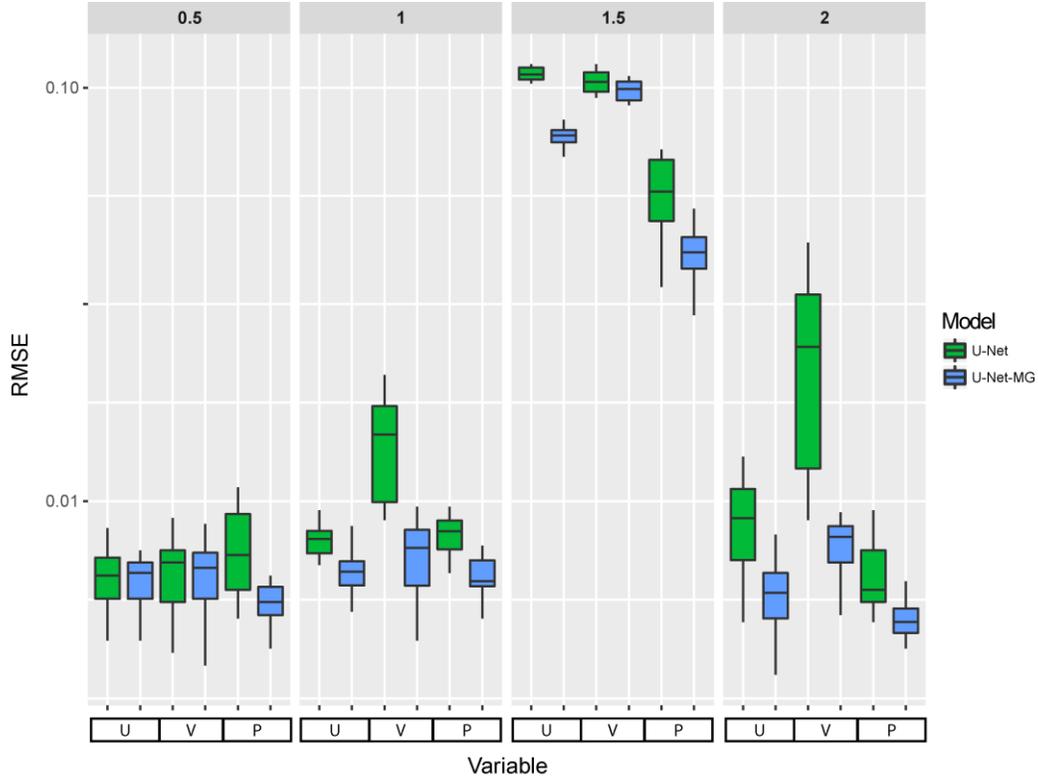

**Fig. 11.** Boxplot of the test RMSE for the U-Net and U-Net-MG models when trained for a set of different inlet velocities. The training dataset is comprised of a subset of snapshots for U = 0.5 m/s, U = 1.0 m/s and U = 2.0 m/s only.

As expected, the test errors are generally higher for extrapolation to a new scenario where no training data was provided (U = 1.5 m/s). However, the errors are generally still under 1% for the scenarios where training data was provided. More importantly, the test errors obtained from the U-Net-MG model are still significantly lower than the U-Net model, with an improvement of 28% for pressure, and a maximum improvement of 34% and 66% for U and V. In particular, the U-Net-MG model is able to extrapolate to the U = 1.5 m/s case with an average reduction in RMSE of ≈ 20% relative to the U-Net model.

### 3.2.3. Hydrodynamics of Oscillating Foil

The U-Net and U-Net-MG models are then trained and tested for the case of flow past an oscillating foil. The input and output fields are similar to the prior case of flow past two cylinders in out-of-phase motion, as illustrated in Fig. 12. However, the instantaneous velocity of the foil is not applied to the foil alone, but to a circle centered on the foil and of diameter 1.5x chord length.

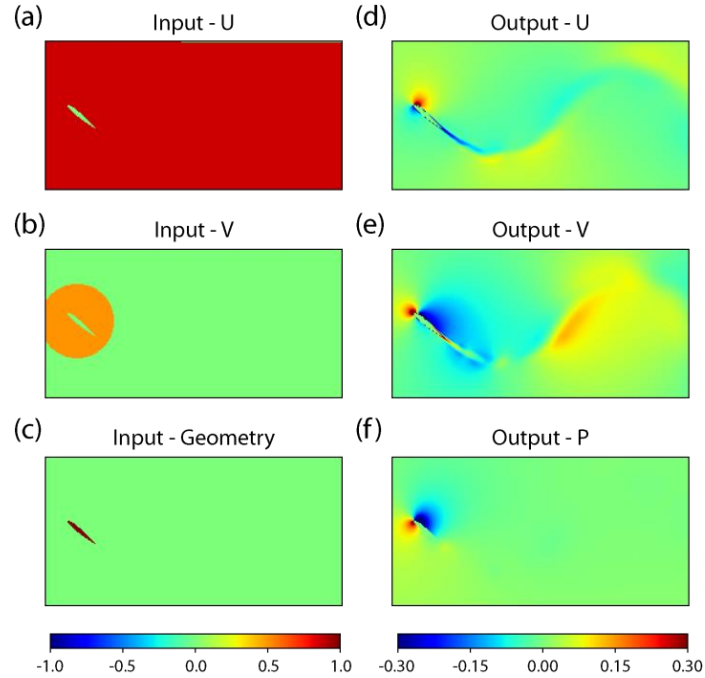

**Fig. 12.** Example of input and output fields provided to the neural network models for training for the case of an oscillating foil. (a), (b) and (c) are input fields encoding U-velocity, V-velocity and a Geometry field, while (d), (e) and (f) are examples of the U-velocity, V-velocity and pressure field as predicted by the neural network.

The NACA 0012 foil creates very different flow patterns under the coupled motions of plunge and rotation. Sample contour plots for U, V and p are provided in Fig. 13, showing how the prediction of V, particularly in the wake region far behind the foil, is much better from the U-Net-MG model than the U-Net model. The difference in accuracy of prediction is further illustrated in Fig. 13 with graphs of U, V, p along the domain through the NACA foil location. In general, the U-Net-MG model performs significantly better for all three parameters (U, V, p) although both networks deviate from the CFD solution in the immediate vicinity of the foil due to the extremely complicated local flow field.

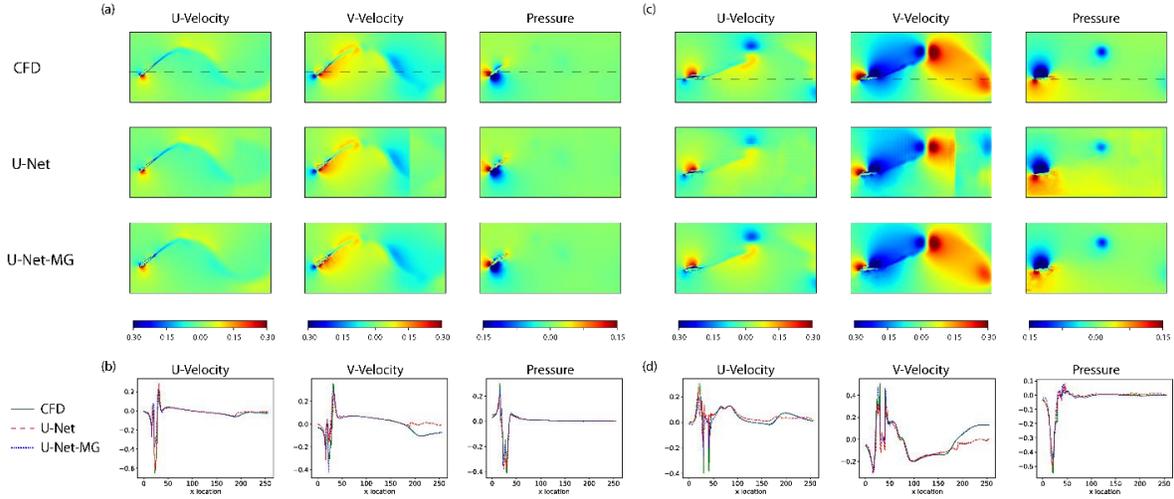

**Fig. 13**. Sample contour plots of flow fields from CFD, U-Net and U-Net-MG for the oscillating NACA 0012 foil under (a) energy harvesting mode, and (c) propulsion mode. (b) and (d) are line graphs showing the values of U, V and P through the domain along the dotted lines depicted in (a) and (c).

The test RMSEs for the two operating modes for NACA 0012 are plotted in Fig. 14, and we demonstrate again that the use of the U-Net-MG architecture can improve the predictive performance relative to the U-Net model, although the error remains around 1%, and is higher than the flow past cylinder examples. Among the three flow parameters, the error in prediction of V is typically the highest. Similar to the previous cases, the use of a U-Net-MG model improves test RMSE significantly from between 36% to 84% across the different parameters.

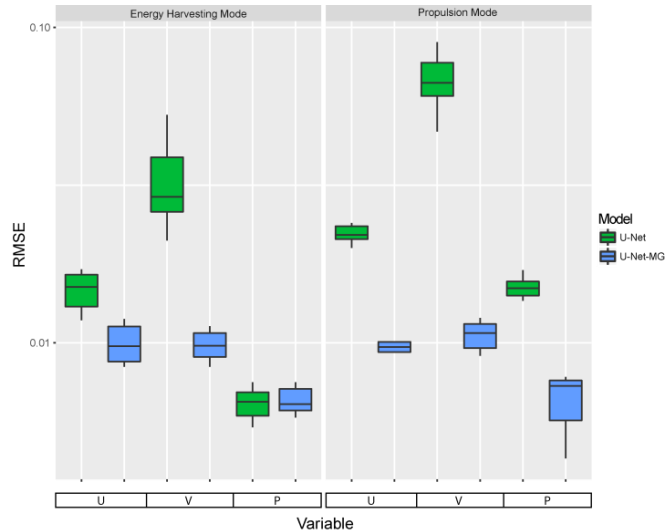

**Fig. 14.** Boxplot of the test RMSE for the U-Net and U-Net-MG models created for 2 flow scenarios of oscillating NACA 0012 foil with a pitch angle of 15º (Propulsion) and 40º (Energy Harvesting)

In addition, a meta-model is trained with the U-Net and U-Net-MG architecture for different NACA geometries. In this case, the U-Net and U-Net-MG models are trained with a subset of simulation data for the NACA 0012, NACA 0020 and NACA 0030 geometries, and tested across four NACA geometries, including the NACA 0025 scenario which was not in the training dataset. Similarly to the set of experiments with variable inlet velocity for flow past two cylinders in out-of-phase motion, the training set here comprises 90 samples of NACA 0012, 75 samples of NACA 0020 and 75 samples of NACA 0030, while the test set comprises 10 samples of NACA 0012, 25 samples of NACA 0020 and NACA 0030, and 100 samples of NACA 0025.

U, V and p fields are predicted for all geometries, and test RMSEs are collated and plotted in Fig. 15. Consistent with our study of a meta-model for different inlet velocities for flow past two cylinders in out-of-phase motion, we note that the test errors for geometries that were not in the training set (NACA 0025) are still lower for the U-Net-MG model than the U-Net model, further illustrating that this proposed architecture can also perform better in extrapolating to other geometries for fluid dynamical systems. In particular, the errors for V, which are typically the highest across the different geometries, were improved by 30% for the case of NACA 0025, which was not in the training set, and improved by up to 70% across the other geometries which were in the training set.

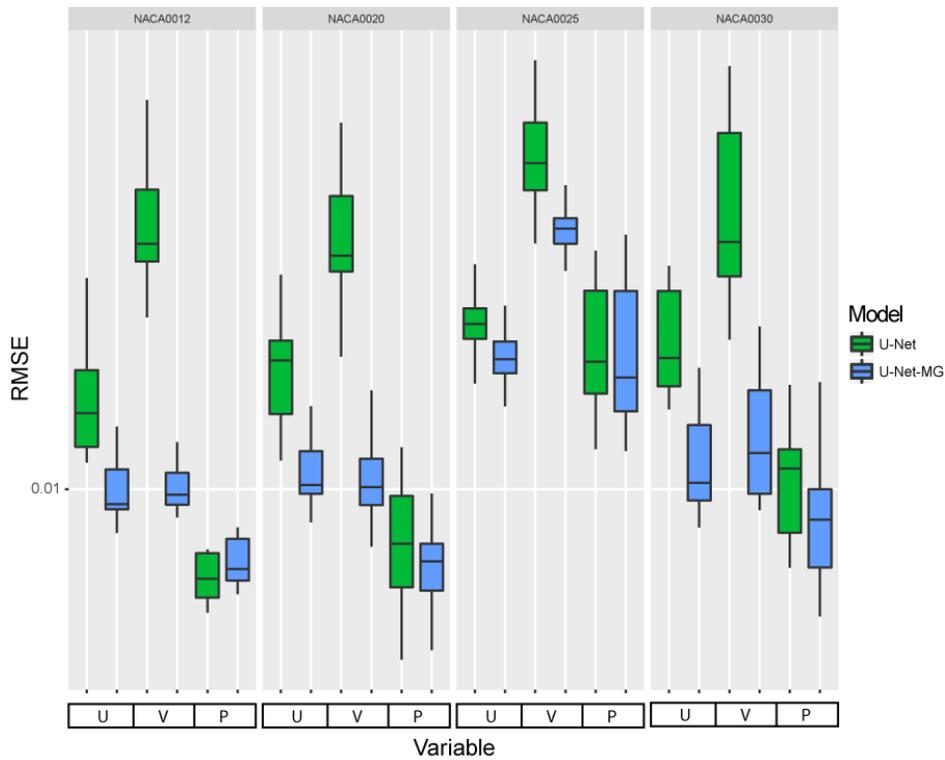

**Fig. 15.** Boxplot of the test RMSE for the U-Net and U-Net-MG models created for a set of different NACA foils under energy harvesting mode (with a consistent $\boldsymbol{\theta_0 = 40^o}$). The NACA 0025 geometry was not part of the training dataset.

## 4. Discussion

In this work, we chose to focus on a set of canonical fluid dynamic systems, as a demonstration of the utility of neural networks in providing fast surrogate models for engineering systems. In particular, although all the cases chosen involve turbulent flow and complex flow physics, the prediction from the models used generally remain under 1% in RMSE, while being much less computationally intensive. Nonetheless, errors persist near the walls and other high gradient regions in the domain where the flow field changes most drastically, and these are potentially important areas to focus on for improvement in future work.

More interestingly, building on the known effectiveness of multigrid numerical methods in CFD and the multi-scale nature of such fluid dynamical systems, we proposed the integration of a multigrid-like structure into the U-Net architecture, and showed that it significantly improves the ability of the neural network to model flow and pressure fields across different flow scenarios and geometrical shapes. Similarly to the fundamental idea behind multigrid solvers, this improvement in representation by the neural network is likely due to the ability of coarser grids to facilitate information transfer between distant nodes on a finer grid more efficiently. In general, across our scenarios studied, we noted an improvement in predictive performance of between 20% and 70% by utilizing the U-Net-MG model over the U-Net model, including for extrapolation to different inlet velocities or different geometries that were not in the original training dataset. This further suggests that any performance improvement observed is not due to overfitting. In particular, the U-Net-MG models learned the variation across different foil shapes better than the conventional U-Net model for the oscillating foil example, suggesting it might have utility in other areas such as the digital twin or design of tidal turbines.

While this is a proof-of-concept of the utility of incorporating the multigrid idea into the U-Net network for physics-based problems, we hypothesize that the U-Net-MG can also be effective when incorporated into other architectures such as variants of generative models like tempo-GAN or RNN-GAN (Junhyuk Kim & Lee, 2020; Xie, Franz, Chu, & Thuerey, 2018), or hybrid spatio-temporal models incorporating CNNs with LSTM or other recurrent neural network models (Han et al., 2019; Hasegawa et al., 2020). We note that generalization to other architectures is likely as previously described in work by He et al.(He & Xu, 2019), while the same principles which suggest the enhancement in performance of CNN-based models via incorporation of multigrid considerations should extend beyond the fluid dynamics applications demonstrated in this work. Hence, there is also potential for exploration and incorporation of such ideas to other engineering applications in future work.

Lastly, we note that this is yet another example to further illustrate the benefits of taking a physics-guided approach to the application of machine learning in engineering systems. Essentially, the choice and appropriate design of a neural network architecture to best match known physical characteristics of the problem is expected to lead to benefits in performance, as illustrated in this work. In particular, corollaries to numerous ideas from the long-established field of scientific computing, such as the multigrid methods in this instance, can be expected to contribute interesting insights in the area of machine learning, and will be an interesting direction to explore in future work.

## 5. Conclusion

The integration of a multigrid-like structure into the U-Net improves the prediction accuracy of the neural network. In particular, we demonstrate an improvement in test RMSEs via the use of a U-Net-MG model architecture over the conventional U-Net architecture for the canonical fluid dynamic cases of flow past a stationary cylinder, flow past two cylinders in out-of-phase motion, and the more challenging and practical case of flow past an oscillating foil. The integration of concepts from the field of scientific computing is expected to yield further benefits in the modeling of engineering systems and fluid dynamic systems, and will be an interesting area to develop in future work.

## Acknowledgments

This study is supported by research funding from the Agency for Science, Technology and Research (A*STAR), Singapore under Grant No. A1820g0084.